\documentclass[%
reprint,
superscriptaddress,
amsmath,amssymb,
aps,
]{revtex4-2}
\usepackage{graphicx}
\usepackage{dcolumn}
\usepackage{bm}
\usepackage{soul}

\begin{document}
	\preprint{APS/123-QED}
	\title{Ultrafast transport and energy relaxation of hot electrons in Au/Fe/MgO(001) heterostructures analyzed by linear time-resolved photoelectron spectroscopy}

\author{Florian K\"{u}hne}
\affiliation{Faculty of Physics and Center for Nanointegration (CENIDE), University of Duisburg-Essen, Lotharstr.~1, 47057 Duisburg, Germany}

\author{Yasin Beyazit}
\affiliation{Faculty of Physics and Center for Nanointegration (CENIDE), University of Duisburg-Essen, Lotharstr.~1, 47057 Duisburg, Germany}

\author{Bj\"{o}rn Sothmann}
\affiliation{Faculty of Physics and Center for Nanointegration (CENIDE), University of Duisburg-Essen, Lotharstr.~1, 47057 Duisburg, Germany}

\author{J. Jayabalan}
\affiliation{Faculty of Physics and Center for Nanointegration (CENIDE), University of Duisburg-Essen, Lotharstr.~1, 47057 Duisburg, Germany}

\author{Detlef Diesing}
\affiliation{Faculty of Chemistry, University of Duisburg-Essen, Universit\"{a}tsstr.~5, 45711 Essen, Germany}

\author{Ping Zhou}
\affiliation{Faculty of Physics and Center for Nanointegration (CENIDE), University of Duisburg-Essen, Lotharstr.~1, 47057 Duisburg, Germany}

\author{Uwe Bovensiepen}
\email[]{uwe.bovensiepen@uni-due.de}
\affiliation{Faculty of Physics and Center for Nanointegration (CENIDE), University of Duisburg-Essen, Lotharstr.~1, 47057 Duisburg, Germany}

\date{\today}

\begin{abstract}
In condensed matter, scattering processes determine the transport of charge carriers. In case of heterostructures, interfaces determine many dynamic properties like charge transfer and transport and spin current dynamics. Here, we discuss optically excited electron dynamics and their propagation across a lattice-matched, metal-metal interface of single crystal quality. Using femtosecond time-resolved linear photoelectron spectroscopy upon optically pumping different constituents of the heterostructure, we establish a technique which probes the electron propagation and its energy relaxation simultaneously. In our approach, a near-infrared pump pulse excites electrons directly either in the Au layer or in the Fe layer of epitaxial Au/Fe/MgO(001) heterostructures while the transient photoemission spectrum is measured by an ultraviolet probe pulse on the Au surface. Upon femtosecond laser excitation, we analyze the relative changes in the electron distribution close to the Fermi energy and assign characteristic features of the time-dependent electron distribution to transport of hot and non-thermalized electrons from the Fe layer to the Au surface and vice versa. From the measured transient electron distribution, we determine the excess energy  which we compare with a calculation based on the two-temperature model that takes diffusive electron transport into account. On this basis, we identify a transition from a super-diffusive to a diffusive transport regime to occur for a Au layer thickness of 20--30~nm.
\end{abstract}
\maketitle
%

%
%
%
%

\section{Introduction}
The propagation of electric currents in metals and semiconductors is on a microscopic level determined by scattering of charge carriers with themselves and with phonons. At interfaces the electronic structure changes abruptly and the necessary energy and/or momentum transfer of charge carriers is mediated by interaction with secondary charge carriers and/or phonons. In case of pseudomorphic interfaces without defects, hybrid electronic wavefunctions develop which conserve energy and momentum at selected points in the electronic band structure $E(\textbf{k})$ across the interface. In the more general case with defects, electron and momentum changes across the interface are compensated by inelastic and elastic scattering processes, respectively. Therefore, charge injection and charge carrier multiplication across interfaces is an interesting problem which is relevant in various energy conversion applications~\cite{hagfeldt_graetzel}. In case of spin-polarized currents generated by, e.g., charge carriers excited in ferro- or ferrimagnetically emitters spin-dependent dynamic properties at such interfaces like, e.g., spin filter effects occur~\cite{hellmann_RMP}.

Analysis of the femto- to picosecond dynamics of optically excited, hot charge carriers in condensed matter provides microscopic information on the interaction processes which these charge carriers experience in the relaxation process~\cite{Shah99,chulkov_ChemRev_2006,Sung2020}. To achieve such a quantitative understanding, it is essential to distinguish transport effects, which spatially redistribute the excited electron density leading to transient inhomogeneous situations, and spatially homogeneous relaxation.

Transport phenomena arise from driving forces such as gradients in occupation number. A gradient in the hot electron density which can be excited by optical absorption of femtosecond laser pulses within the optical absorption depth drives electron currents which can be either ballistic or diffusive depending on the ratio between scattering length and sample size. A gradient in temperature results in heat flow~\cite{Hohlfeld2000,liu_05}. Such transport effects have been observed in early studies~\cite{Brorson1987}, and were empirically~\cite{Hohlfeld2000} as well as microscopically~\cite{Battiato2012} taken into account in theoretical modeling. While these effects are particularly relevant in surface sensitive experiments like, e.g. time-resolved photoemission~\cite{aeschlimann_APA00,liso_APA04,liso_APA04b,klick_2019} and surface second harmonic generation~\cite{Hohlfeld2000} they also facilitated separation of transport from relaxation effects in linear optical experiments due to differences in depth sensitivity of the real and imaginary part of the optical response~\cite{wieczorek_2015}.

Brorson \textit{et al.}~\cite{Brorson1987} have established a direct experimental approach to distinguish transport (non-local effects) from relaxation (local effects) in pump-probe experiments by pumping and probing at opposite or identical sides of the sample of interest, respectively. The resulting transient electronic population can modify chemical bonding of molecules on surfaces in catalytic surface reactions~\cite{bonn_PRB00,liso_APA04}. The back-side excitation geometry allows to pump the system exclusively by hot electrons contrary to optical excitations which involve initial state in, e.g., HOMO-LUMO transitions of adsorbed molecules. More recently, this backside pump approach was used by Melnikov \textit{et al.} to investigate spin currents in Fe/Au heterostructures~\cite{Melnikov2011,Razdolski2017a,Alekhin2017,Alekhin_2019} by linear and nonlinear optical pump-probe experiments and by Bergeard \textit{et al.}~\cite{bergeard_2016} to analyze hot electron induced demagnetization of Co/Pt heterostructures by linear optical means. In a more recent work, ballistic charge carrier propagation has been investigated in such a configuration in perovskite thin films~\cite{Sung2020} where it provided important insights into energy-dependent scattering processes. Such spectroscopy was demonstrated in time-resolved two-photon photoemission experiments on epitaxial heterostructures Au/Fe/MgO(001) in which the pump pulse excited the Fe and the photoelectrons were emitted from the Au surface~\cite{beyazit_2020}. In this work, we demonstrated that achieving ballistic currents in these samples is challenging because the time scale for ballistic propagation through the sample at the Fermi velocity is close to the average electron-electron scattering time which were reported earlier~\cite{chulkov_ChemRev_2006,bauer_2015}. In such a case, the electronic transport is considered to be super-diffusive rather than ballistic~\cite{Battiato2012}.
Moreover, the spin-dependent dynamics at such heterostructures consisting out of ferromagnetic metals and heavy metals is known as THz emitters due to the spin-dependent currents across these interfaces~\cite{kampfrath_2013}. Understanding the spatio-temporal electron distribution might facilitate a microscopic understanding in these emission processes and help to distinguish spin-polarized charge vs. magnon currents.

In this article, we report our results on time-resolved linear photoemission spectroscopy ({\it tr}-PES) performed on epitaxial Au/Fe/MgO(001) heterostructures of different Au thicknesses. In these measurements, the photoemission spectrum of the Au layer in the vicinity of the Fermi energy ($E_{\mathrm{F}}$) was probed while carriers were excited either directly in the same Au layer or in the nearby Fe layer. We analyze the dependence of the hot-electron dynamics as a function of the Au layer thickness and identify signatures of electronic transport in the non-equilibrium electron distribution function, which we discuss on the basis of the time-dependent excess energy.

\section{Experimental Details}\label{Sec:Exp}

\begin{figure}
\centering
\includegraphics[width=1.0\columnwidth]{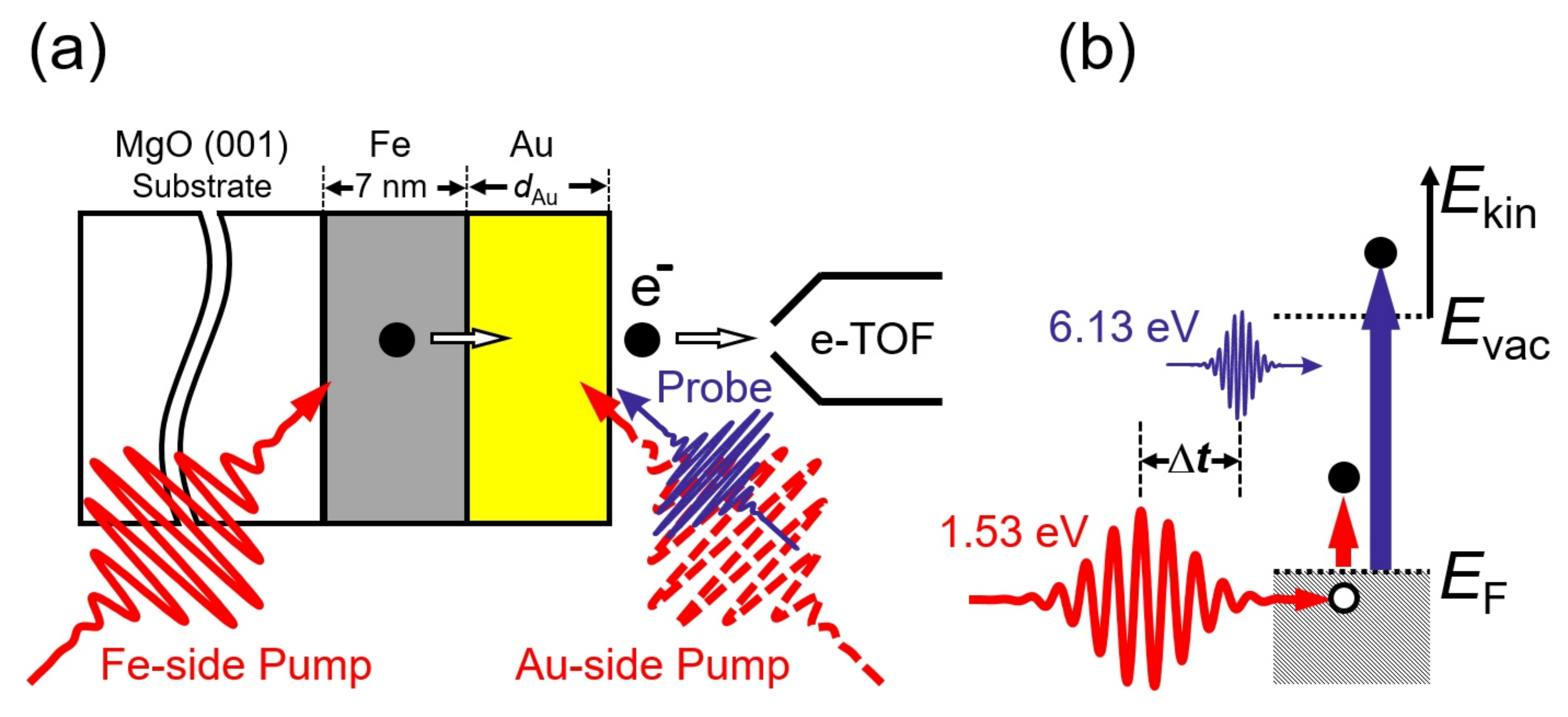}
\caption{(a) Schematic of the sample configuration along with the two pump and probe configurations used in the measurement of the $tr$-PES with an electron time of flight spectrometer (e-TOF). A 1.53~eV infrared pump beam excites carriers directly either in the Au layer (Au-side) or in the Fe layer (Fe-side) while a 6.13~eV UV pulse always probes the Au surface. (b) Energy diagram of the $tr$-PES measurement with a time delay $\Delta t$ between the pump and probe pulses, $E_{\mathrm{vac}}$ is the vacuum energy, $E_{\mathrm{F}}$ is the Fermi energy of the metallic heterostructure and $E_{\mathrm{kin}}$ the kinetic energy of the photoelectrons.}
\label{Fig:SchematicLinearPESAuFeMgO}
\end{figure}

The samples under study are epitaxial Au-Fe heterostructures grown on a transparent MgO(001) substrate using molecular beam epitaxy. A schematic of the sample is shown in Fig.~\ref{Fig:SchematicLinearPESAuFeMgO}(a). In a preparation chamber, Fe(001) was grown on the MgO(001) substrate which was followed by growth of Au(001). Note that the in-plane axes of both layers are rotated by $\pi/4$ with respect to each other to minimize the lattice mismatch between Fe and Au and facilitate pseudomorphic growth~\cite{Muehge_APA94, Mattern_22}. The MgO allows for an almost transparent path for direct optical excitation of the Fe layer in the pump-probe measurements. The thickness of the layers were determined by AFM and depth analysis of grooves through the whole film stack which were prepared by a needle. In this work, three different Au layers of thickness $d_{\mathrm{Au}} = 5$~nm, 15~nm, and 28~nm were investigated. The Fe layer was kept at a fixed thickness  $d_{\mathrm{Fe}} = 7$~nm. The time-resolved photoelectron spectroscopy measurements were carried out after sample transfer through ambient conditions into the vacuum chamber equipped with the photoelectron spectrometer under ultrahigh vacuum conditions in two different pump configurations, see Ref.~\cite{SANDHOFER2014} for a detailed description of the photoemission setup. The schematic of the pump and probe configurations are shown Fig.~\ref{Fig:SchematicLinearPESAuFeMgO}(a). In the ``Au-side'' configuration, both the pump and probe pulses arrive directly on the Au surface. In the ``Fe-side'' configuration, the pump pulse reaches the Fe layer by entering through MgO substrate while the probing is still done on the Au surface, spatially separating the electron excitation from the probe at the surface. The infrared pump pulses at 1.53~eV used in the {\it tr}-PES measurements were obtained directly from the output of a Ti:Sapphire amplifier (Coherent RegA 9040) operating at a repetition rate of 250~kHz. The spectral width of the pump pulse was 70~meV full width at half maximum (FWHM). The 6.13~eV probe pulses were obtained by generating the forth harmonic of the fundamental 1.53~eV beam using two consecutive second-harmonic generations in a Beta barium borate crystal ($\beta$-BBO). Both the pump and probe beams were $p$-polarized. The angle of incidence of both the pump and probe beams on the sample were close to 45$^{\circ}$. In the Fe-side pump configuration, the pump beam has an incidence angle of 45$^{\circ}$ towards the sample and 90$^{\circ}$ with respect to the probe beam. The pump fluence used in all the measurements was about 100~$\mu$Jcm$^{-2}$. It was limited by photoelectron emission due to multiphoton absorption of the pump pulse. The typical full-width half-maximum (FWHM) of the pump beam was usually between $100-150~\mu$m with the probe beam focus being roughly $30\%$ smaller to ensure homogeneous excitation. The kinetic energy of the photoelectrons emitted by the sample within an angle $\pm 11^{\circ}$ to the surface normal was collected and measured using a time-of-flight electron spectrometer (e-TOF). The photoemission process in the Au layer is illustrated by Fig.~\ref{Fig:SchematicLinearPESAuFeMgO}(b). All the measurements were carried out at room temperature. The time resolution of the setup was determined by measuring the temporal width of the highest 100~meV electrons in the correlated photoemission signal which was found to be below 100~fs.

\begin{figure}
\centering
\includegraphics[width=0.8\columnwidth]{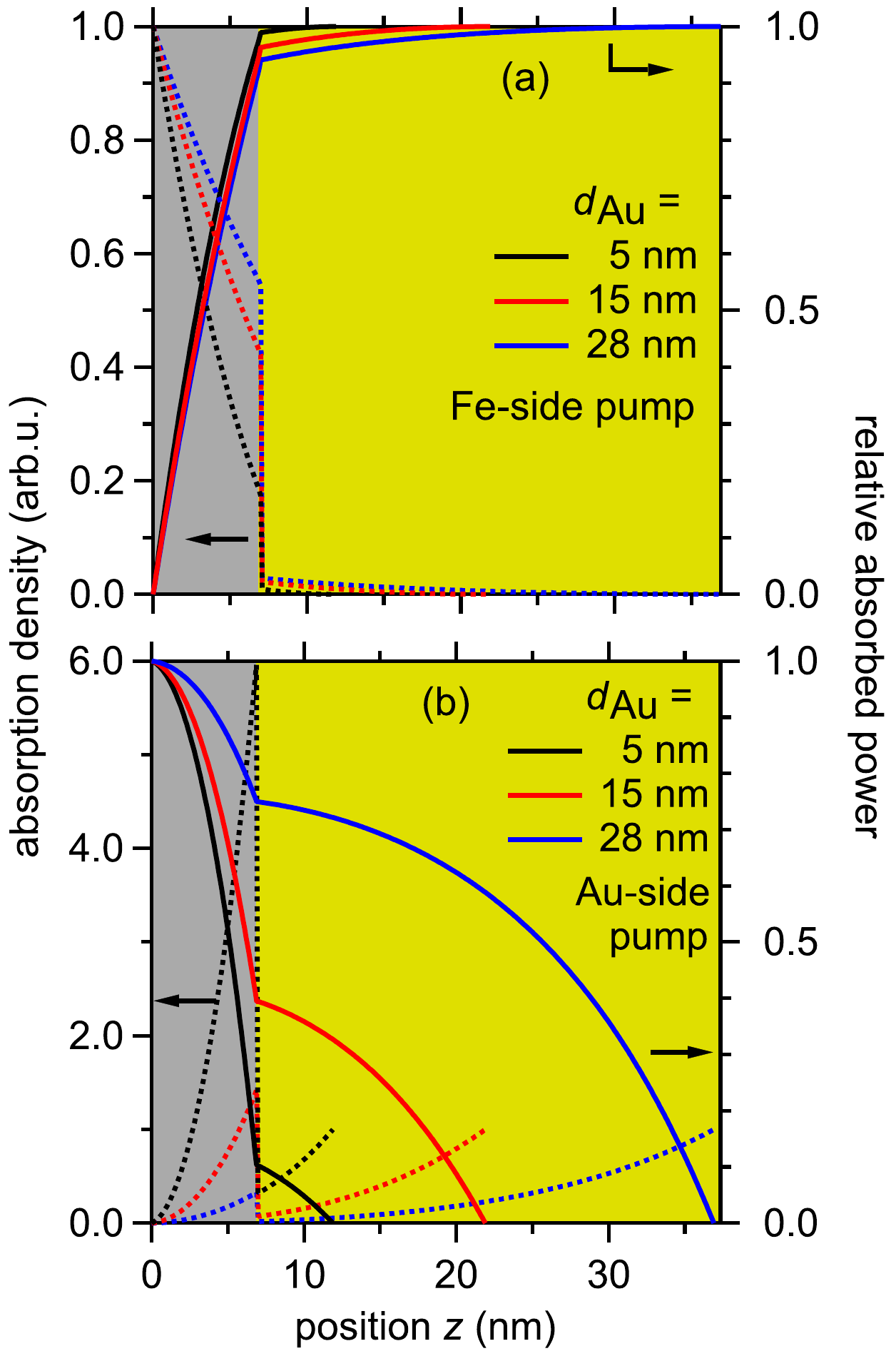}
\caption{The calculated relative pump light intensity with respect to the incident light field (dashed lines, left axis) and the relative absorbed pump light intensity (solid lines, right axis) as a function of Au layer thickness for (a) Fe-side and (b) Au-side pumping. The Fe layer thickness is fixed at 7 nm. The maximum in absorption density in (b) is higher at the interface because of the change in optical constants.}
\label{Fig:Abs}
\end{figure}

To understand to which degree absorption of the pump pulse occurs in the Au and Fe layers, we calculate the electric field inside the material in both pump configurations using the IMD software~\cite{windt_imd} and derive the absorbed power $P(z)$ in the different constituents
\begin{equation}
P(z) = n(z)I(z) = n(z)|E(z)|^2 = n(z)I_{t}e^{-\alpha z},
\label{eq:2}
\end{equation}
with $z$ being the distance from the Fe-Au interface along the normal direction, $\alpha$ the absorption coefficient, $I(z)$ the intensity, $E(z)$ the electric field and $n(z)$ the refractive index of the material. The power in the layer stack is given by the real part of the Poynting vector. Fig.~\ref{Fig:Abs} depicts the relative intensities for (a) Au-side and (b) Fe-side pumping, as well as the relative absorbed power, which we obtained by subtracting the transmitted field and normalizing the incoming intensity to 1 because we are only interested in the attenuation of the field by absorption. To determine the spatial distribution of the optical absorption in the heterostructure, we used the pump photon energy $E = 1.53~\mathrm{eV}$, the angle of incidence $\theta = 45^{\circ}$, the $p$-polarization of the light, the refractive indices $n_{\mathrm{Au}}$(1.53~eV) = 0.08, $n_{\mathrm{Fe}}$(1.53~eV) = 3.02 and the extinction coefficients $k_{\mathrm{Au}}$(1.53~eV) = 4.69 $k_{\mathrm{Fe}}$(1.53~eV) = 3.72~\cite{weaver81}. We find that in case of Fe-side pumping the absorption is almost exclusively occurring inside the Fe layer: 99\% for $d_{\mathrm{Au}} = 5$~nm and 94\% for $d_{\mathrm{Au}} = 28$~nm. The case for Au-side pump is more involved. In the lower panel of Fig.~\ref{Fig:Abs} we recognize a strong variation in the relative intensity at the Fe-Au interface because of the change in refractive index and the light field is attenuated in Au when it reaches Fe, the more the thicker the Au layer is. The intensity that reaches Fe decreases with $d_{\mathrm{Au}}$.  At $d_{\mathrm{Au}} = 5$~nm,  11\% of the pump is absorbed in Au, while the majority is absorbed in the Fe layer. For $d_{\mathrm{Au}}$ = 28 nm, the absorption of the pump mostly occurs in the Au layer and for $d_{\mathrm{Au}}$ = 15~nm around 40\% is absorbed in Au and 60\% in Fe. We note that a systematic comparison of the pump-induced dynamics as a function of $d_{\mathrm{Au}}$ should be done for comparable pump conditions, i.e. Fe-side pumping. Au-side pumping distributes the pump energy in a non-trivial manner across the layer stack.

\begin{figure}
\centering
\includegraphics[width=0.8\columnwidth]{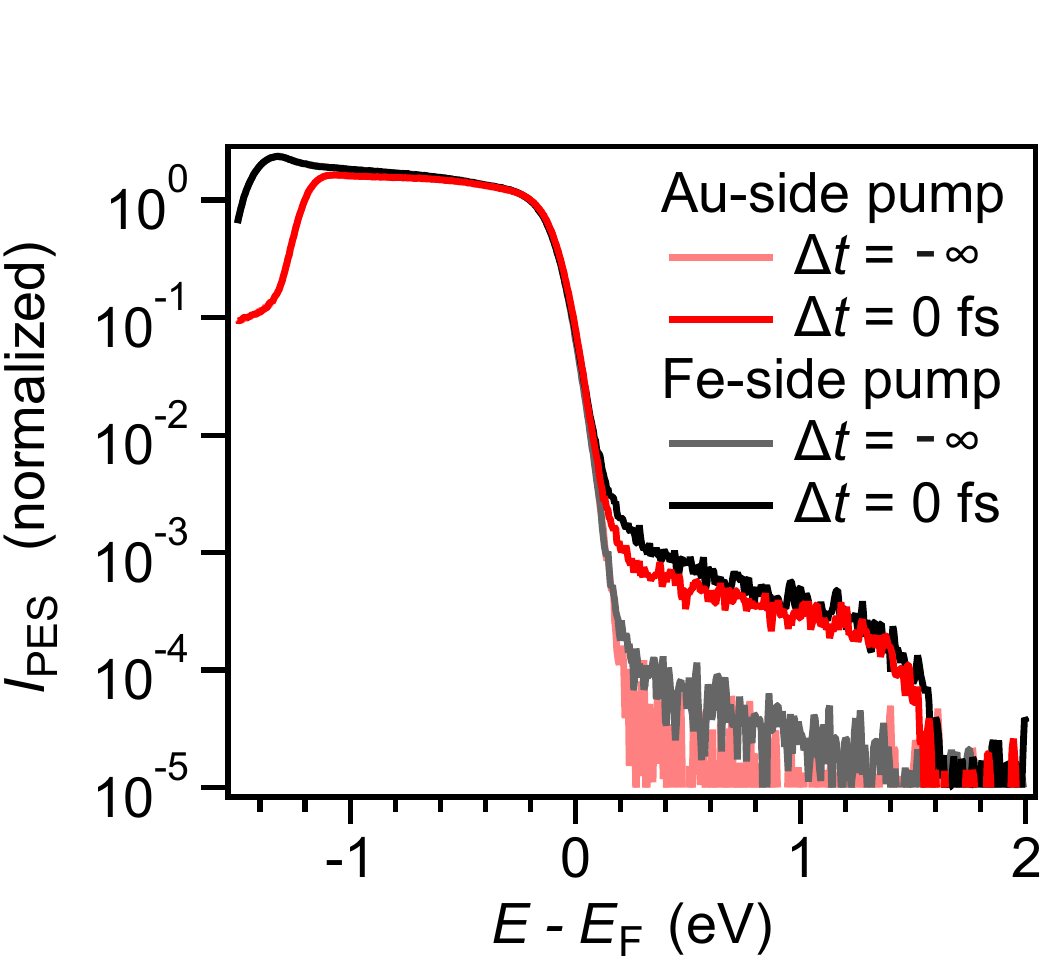}
\caption{Photoemission intensity $I_{\mathrm{PES}}(E)$ measured on the Au surface of the Au/Fe/MgO(001) heterostructure with $d_{\mathrm{Au}} = 5$~nm at two different time delays for Fe-side (black) and Au-side (red) excitations. The photoemission spectra measured much before the arrival of the pump pulse are shown in light colors while the spectra obtained at time zero are shown in full colors. The spectra are normalized at $E-E_{\mathrm{F}} = -0.1$~eV. The incident pump fluence used in the experiment was about $100 \mu$Jcm$^{-2}$}
\label{Fig:PEIVsE}
\end{figure}

\section{Experimental Results}

Figure~\ref{Fig:PEIVsE} shows the recorded photoelectron spectra $I_{\mathrm{PES}}(E)$ well before the pump pulse arrival $\Delta t = - \infty$ which we term $I^0_{\mathrm{PES}}(E)$, and at temporal pump-probe overlap $\Delta t = 0$ for $d_{\mathrm{Au}}=5$~nm. Results for the cases of Fe- and Au-side excitation are depicted. The measured photoemission spectra contain signatures of photoexcited holes at $E-E_{\mathrm{F}}<0$~eV for both configurations. However, in the present manuscript we discuss the changes above $E_{\mathrm{F}}$ due to the better statistics of the data, which allows for a detailed analysis.  For energies above $E_{\mathrm{F}}$, the spectrum at $\Delta t = - \infty$ follows a Fermi-Dirac distribution taking into account spectral broadening due to the ultraviolet femtosecond probe laser pulse with a bandwidth $\Delta E=60$~meV and room temperature, where the data were recorded. The difference in the low-energy cutoff near $E-E_{\mathrm{F}}< -1$~eV which is determined by the workfunction, is explained by the different positions on the Au surface at which the spectra were recorded and local variation in the workfunction. With the absorption of the infrared pump pulse, electrons in the states up to 1.53~eV below $E_{\mathrm{F}}$ are excited to states up to 1.53~eV above $E_{\mathrm{F}}$. One can clearly recognize this absorption by the increase in photoemission intensity relative to before the pump-pulse arrival by an order of magnitude to $10^{-4}$ at $\Delta t$ = 0 in that energy range. Note that the spectra at $\Delta t$ = 0 for Fe- and Au-side pumping are found to be almost identical for the 5 nm thick Au layer. For the further discussion the pump-induced changes in the photoelectron spectra $\Delta I_{\mathrm{PES}}(E,\Delta t)=I_{\mathrm{PES}}(E,\Delta t)-I^0_{\mathrm{PES}}(E)$ are calculated.

\begin{figure}
\centering
\includegraphics[width=1.0\columnwidth]{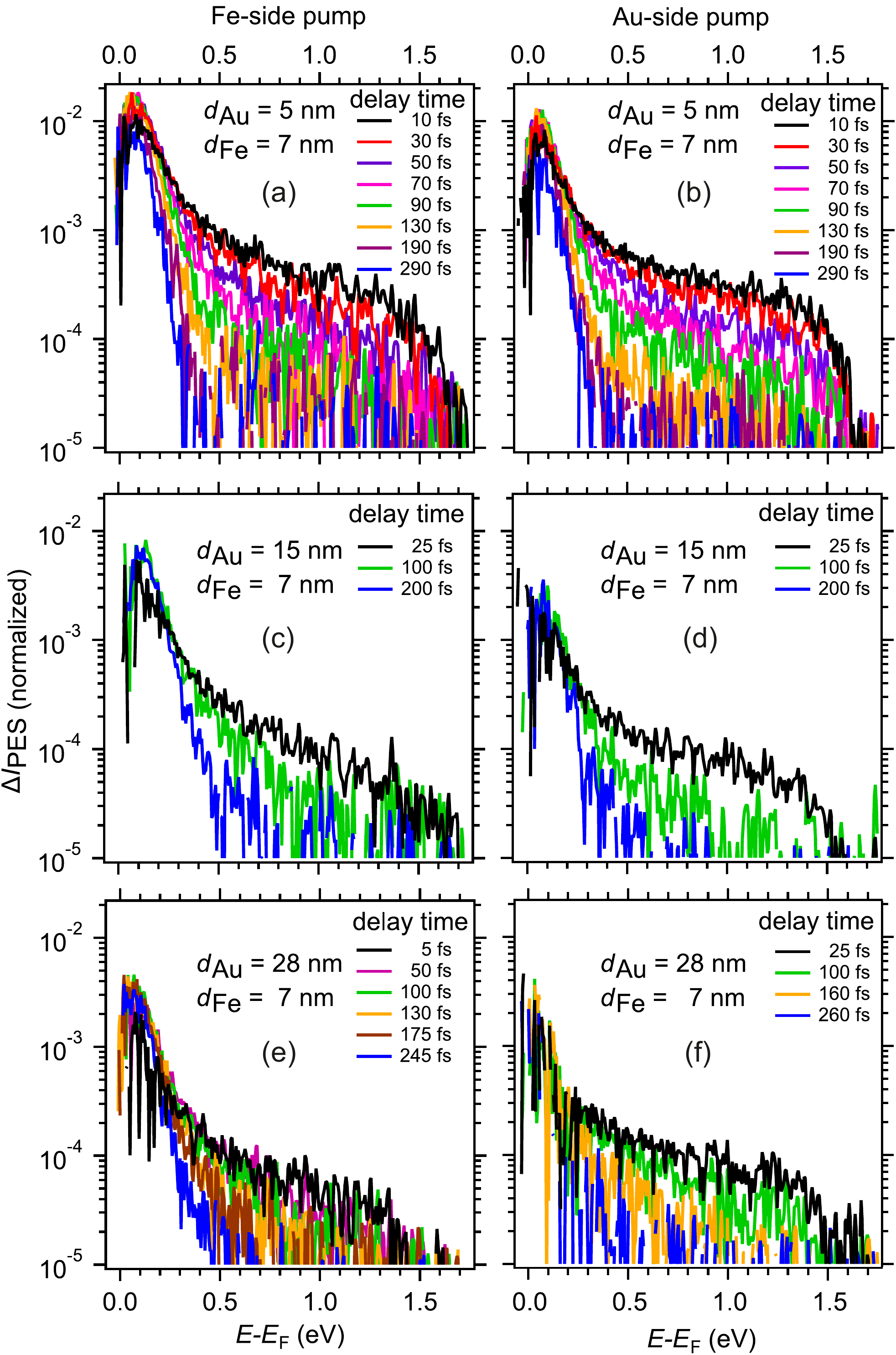}
\caption{Pump-induced changes in the time-resolved photoemission intensity $\Delta I_{\mathrm{PES}}(\Delta t)$ at energies above the Fermi level for Fe-side (a,c,e) and Au-side (b,d,f) pumping for the three different Au layer thicknesses $d_{\mathrm{Au}}$ = 5, 15, and 28~nm. The incident pump fluence was $100~\mu$Jcm$^{-2}$.  Spectra recorded at delay times in between the depicted ones were averaged symmetrically to improve the statistics and the variation in the time delays between the depicted spectra represent the statistics of the original datasets.}
\label{Fig:DelPESVsEfords}
\end{figure}

In Fig.~\ref{Fig:DelPESVsEfords}(a,b), we show $\Delta I_{\mathrm{PES}} (E, \Delta t)$ at selected delay times for Fe- and Au-side excitation of the heterostructure with $d_{\mathrm{Au}} = 5$~nm. Such measurements were also performed for $d_{\mathrm{Au}} = 15, 28$~nm and are also shown in Fig.~\ref{Fig:DelPESVsEfords}. In all the samples and independent of the side of excitation, the absorption of the pump pulse causes significant changes in $\Delta I_{\mathrm{PES}}$ in the energy range of 0 to 1.5~eV at $\Delta t$ = 0. The time-dependent evolution $\Delta I_{\mathrm{PES}}(\Delta t)$ strongly depends on the electron energy and proceeds differently above and below $E$-$E_{\mathrm{F}}$ = 0.5~eV. We distinguish two contributions. At higher energies, the transient electron distribution depends exponentially on energy and relaxes with increasing $\Delta t$. The second contribution occurs for $E-E_{\mathrm{F}} < 0.5$~eV and contains about one order of magnitude more electrons per energy. Further, up to nearly 100~fs the overall distribution is clearly non-thermal, i.e. it deviates from a Fermi-Dirac distribution.

\begin{figure}
\centering
\includegraphics[width=0.8\columnwidth]{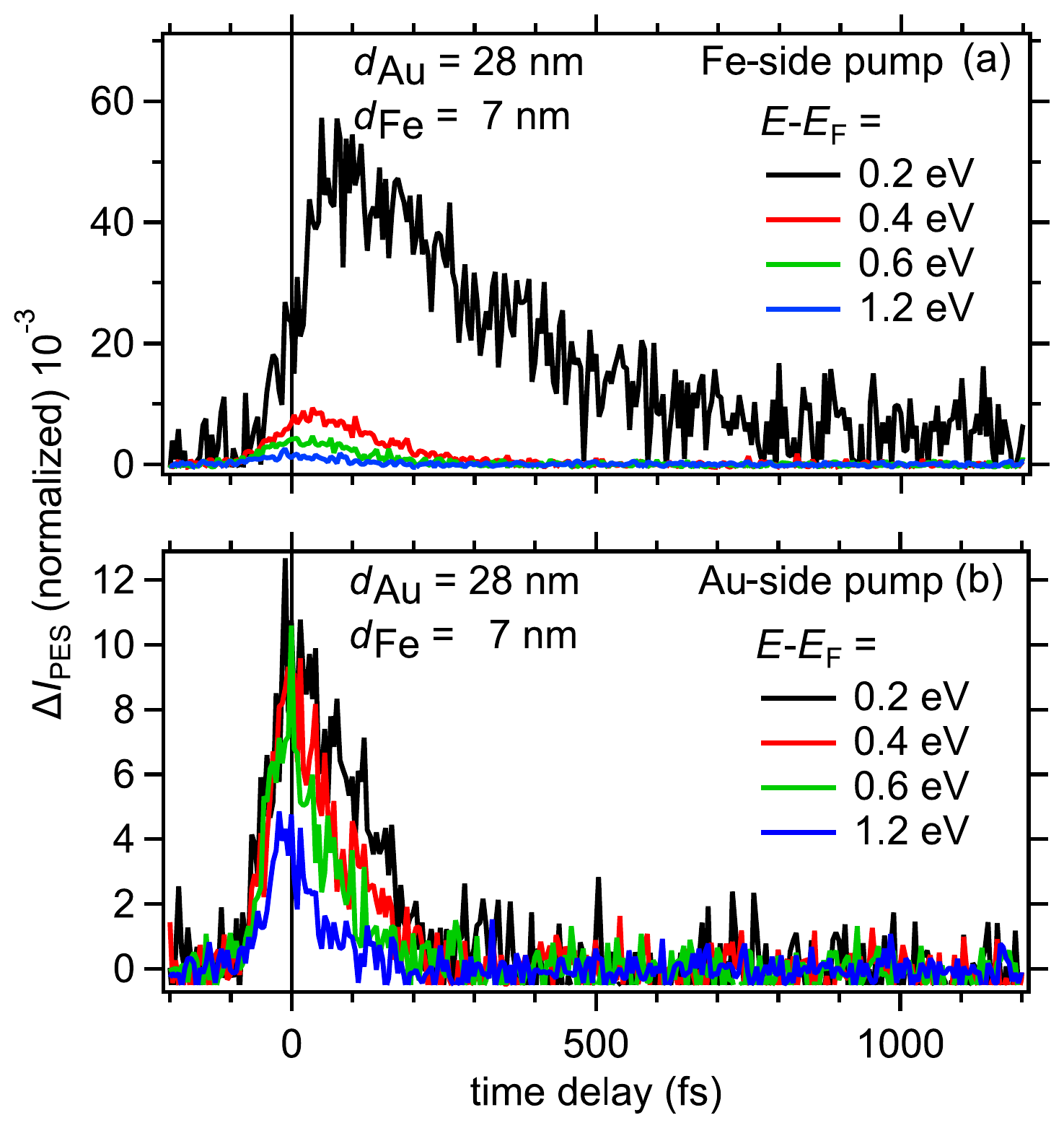}
\caption{Time-dependent change in photoemission intensity $\Delta I_{\mathrm{PES}}(\Delta E)$ upon laser excitation at energies above the Fermi level for a Au layer thickness $d_{\mathrm{Au}}$ of 28~nm in both (a) Fe-side as well as (b) Au-side pumping. The intensities were integrated over an energy $\Delta E$ of 200~meV. Only selected energies in the 1.55 eV high pump induced change are shown for visibility.}
\label{Fig:FigNEWdelay}
\end{figure}

In order to compare the difference in the temporal response at various energies, we plot the time-dependent change in photoemission intensity, $\Delta I_{\mathrm{PES}} (E, \Delta t)$, for selected energies in Fig.~\ref{Fig:FigNEWdelay} for Fe- and Au-side excitations of the heterostructure with $d_{\mathrm{Au}} = 28$~nm and $d_{\mathrm{Fe}} = 7$~nm. We observe that the magnitude of $\Delta I_{\mathrm{PES}}$ of the low energy electrons of the Fe-side pump case is much higher, it increases slowly reaching a maximum at a delay time of 100~fs, relaxes within 800~fs, and has a weak pedestal at longer delays. In the Au-side pump case, Fig.~\ref{Fig:FigNEWdelay} (b), the change in photoemission intensity is much weaker compared to the Fe-side case, even though we are directly probing the pumped region. Further, the maximum $\Delta I_{\mathrm{PES}}$ at all energies are reached almost at zero time delay and they all decay faster. Thus, the behavior of the low energy electrons of the Fe-side case is very different from all others irrespective of the pumping configuration. To investigate this effect in more detail, we calculate the time-dependent energy content at low and high energy regimes from these data.

The energy and time-dependent relative change in the electron population $\Delta n$($E$,$\Delta t$) is assumed to be proportional to the measured $\Delta I_{PES}$. The time-dependent energy content in a given energy range $E_1$ to $E_2$ is then calculated by~\cite{liso_APA04},
\begin{equation}
	\epsilon_{E_1}^{E_2} (\Delta t) = \mathcal{C} \int_{E_1}^{E_2} \Delta n \left(E,\Delta t\right) \left(E-E_{\mathrm{F}}\right) dE,
	\label{Eq:EnergyDensity}
\end{equation}
where $\mathcal{C}$ is a proportionality constant which depends on photoelectron detection efficiency. It is determined by the photoemission matrix element and geometric factors of the setup. Since the probing geometry and photon energy are kept constant for all measurements reported here, the integrals in Eq.~\ref{Eq:EnergyDensity} resulting for the different datasets can be compared among each other. Therefore, $\epsilon_{E_1}^{E_2}$ represents a normalized quantity which provides access to the relative excess energy content.

\begin{figure}
\centering
\includegraphics[width=\columnwidth]{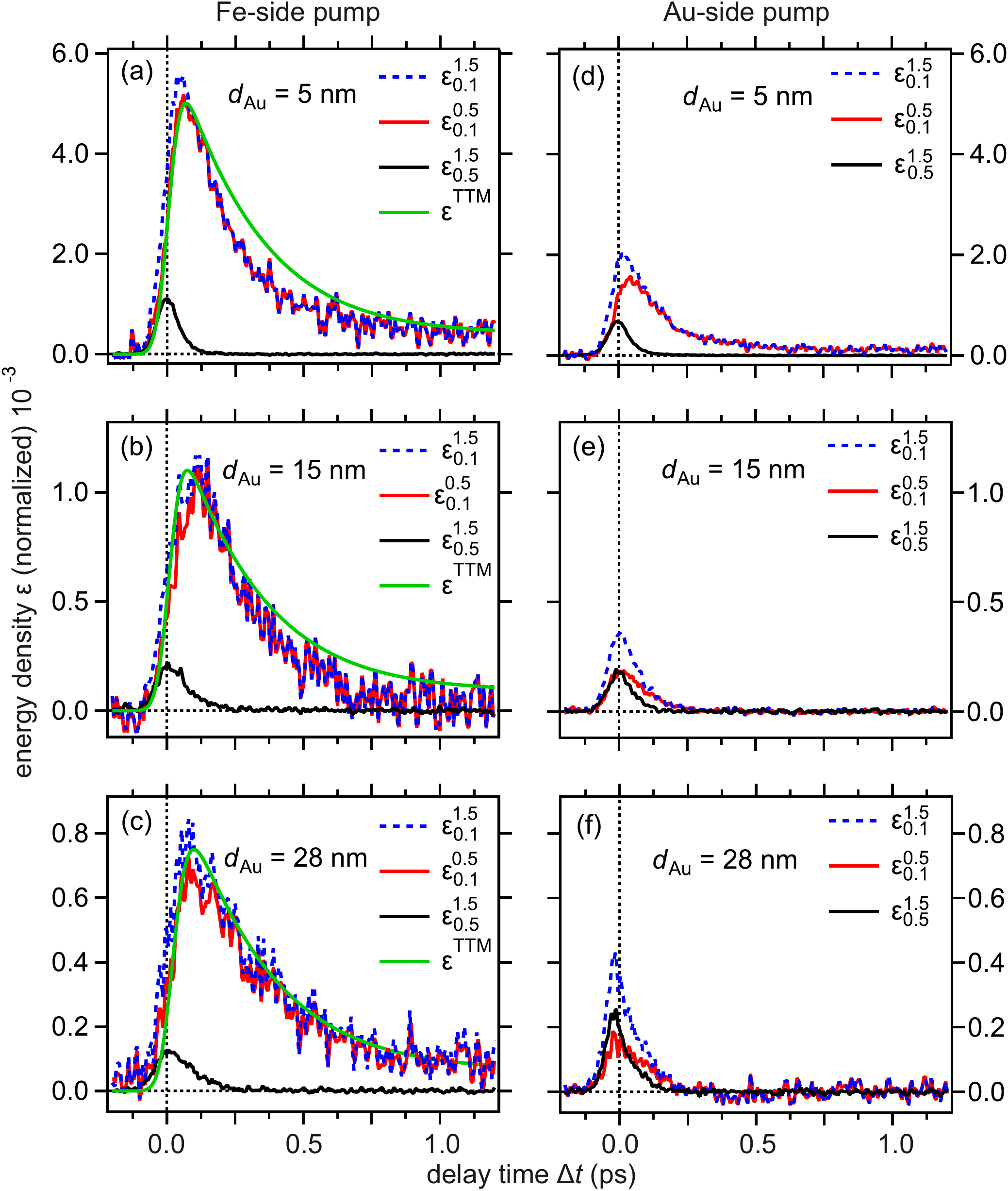}
\caption{The time-dependence of the energy densities $\epsilon$ calculated at different energy ranges using Eq.~\ref{Eq:EnergyDensity} for the Fe-side (a, b, c) and Au-side (d,e,f) excitation. The determined $\epsilon$ can be compared quantitatively for the different experimental datasets. The green traces are energy densities calculated by an adapted two-temperature model (TTM) and convolved with a 80~fs FWHM Gaussian pulse. They are scaled to the experimental data by a factor, see text for details.}
\label{Fig:EdensityVsDeltat}
\end{figure}

In Fig.~\ref{Fig:EdensityVsDeltat}, we show these time-dependent energy densities, the total measured energy density $\epsilon_{0.1}^{1.5}$ in the full energy range (0.1 to 1.5~eV), $\epsilon_{0.1}^{0.5}$ for the lower-energy excited electrons, and $\epsilon_{0.5}^{1.5}$ for the higher-energy excited electrons of all samples and both pump configurations. In all the cases, the $\epsilon_{0.1}^{1.5}$ plotted in Fig.~\ref{Fig:EdensityVsDeltat} shows a rise of the excited carriers at time zero with the arrival of the pump pulse followed by a decay at later times. A comparison of the measured total energy density shows that for one selected pumping geometry the maximum $\epsilon_{0.1}^{1.5}$ reduces with the increase in the Au layer thickness. Further, the maximum $\epsilon_{0.1}^{1.5}$ is always higher for the Fe-side pumping if compared to that of corresponding data for Au-side excitation. These observations are well explained by the absorption calculations discussed earlier in Fig.~\ref{Fig:Abs}. The pump absorption is much larger if electrons are excited in Fe compared with Au-side pumping. Apart from the case of $d_{\mathrm{Au}} = 5$~nm, the decay is slower for the Fe-side excitation when compared to that of Au-side excitation. In case of $d_{\mathrm{Au}} = 5$~nm the decay occurs on identical time scales for both pumping configurations, see Fig.~\ref{Fig:TTM}(d) for a direct comparison. These slower decaying cases always show a delayed maximum in $\epsilon_{0.1}^{0.5}$ with respect to time zero. Further, we find that for all $d_{\mathrm{Au}}$ in the Fe-side pump configuration $\epsilon_{0.5}^{1.5}$ decays faster and is lower in magnitude when compared to that of the corresponding $\epsilon_{0.1}^{0.5}$ (Fig.~\ref{Fig:EdensityVsDeltat}). Thus, after a build-up of low-energy electrons due to scattering of high energy electrons, the low-energy electrons ($<$ 0.5~eV) comprise the majority of the measured energy density.

\section{Two-temperature Model}

\begin{figure}
\centering
\includegraphics[width=0.99\columnwidth]{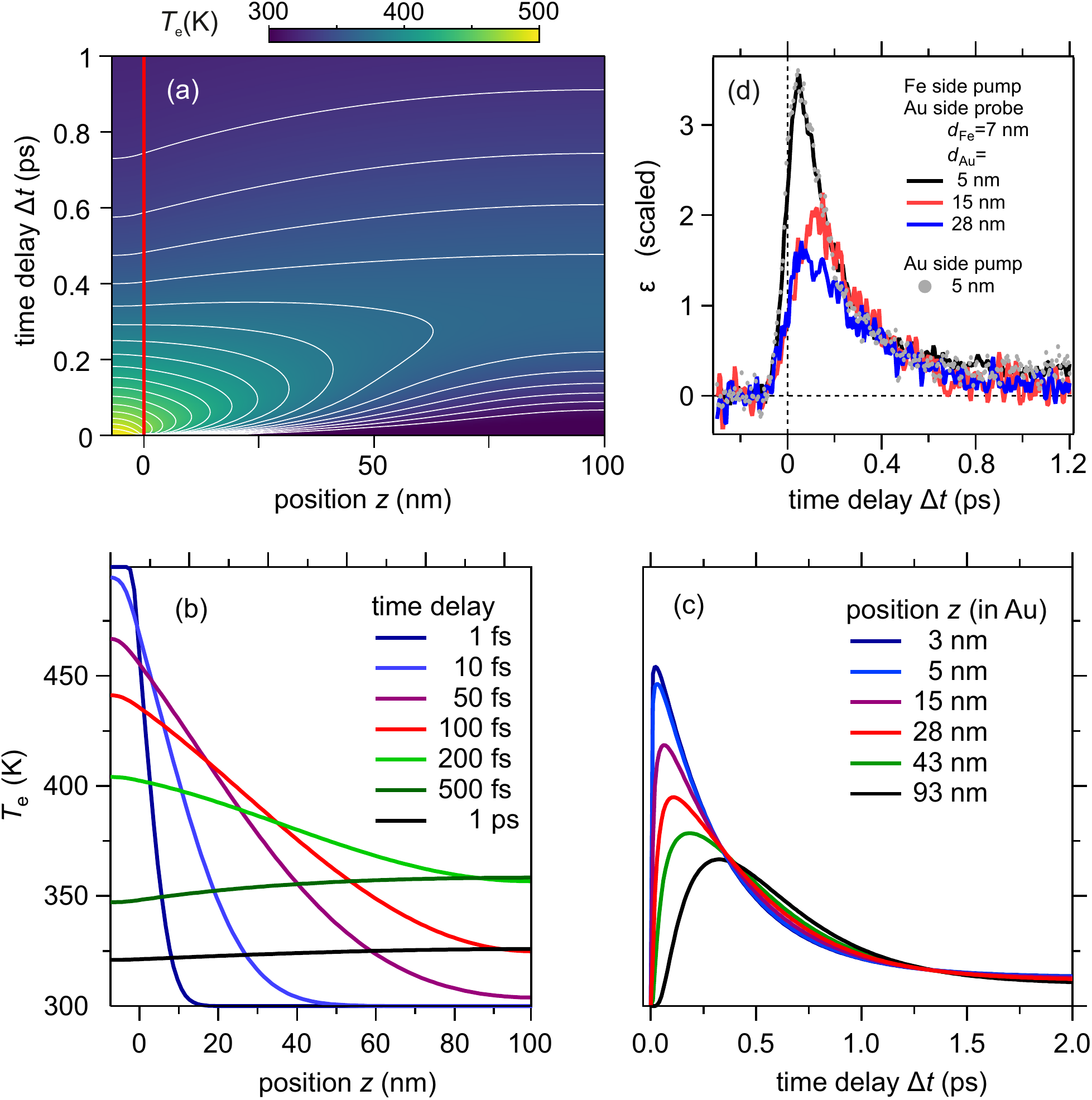}
\caption{(a) Combined false color and contour plot of the calculated electron temperature by using the two-temperature model (TTM) as detailed in the text for $d_{\mathrm{Fe}}$ = 7~nm and $d_{\mathrm{Au}}$ = 100~nm. Two contour lines are separated by 10~K. The horizontal axis describes the position $z$ along the interface normal; $z<0$ represents Fe and $z>0$ the Au layer. The change with $\Delta t$ is plotted along the vertical axis. The electronic heat is generated by an instantaneous increase of $T_e$ in Fe from 300 to 500~K. (b) Temperature transients at indicated values of $z$ as a function of $\Delta t$. $z = 0$ refers to the Fe/Au interface. (c) Time evolution of the electron temperature at selected $z$. (d) Low energy density $\epsilon_{0.1}^{0.5}$ for Fe-side pumping of $d_{\mathrm{Au}}$ = 5, 15, 28~nm and Au-side pumping of $d_{\mathrm{Au}}$ = 5~nm. The datasets are scaled to each other to match at $\Delta t = 0.5$~ps.}
\label{Fig:TTM}
\end{figure}

We compare the observed electron-propagation dynamics in case of Fe-side pumping with model predictions in the limit of thermalized electron distributions as a function of $\Delta t$ and position $z$ along the interface normal direction. To this end, we adopted the two-temperature model (TTM)  by Anisimov {\it et al.}~\cite{anisimov_sovphys74}  by (i) considering Au and Fe films of thicknesses $d_{\mathrm{Au}}$ and $d_{\mathrm{Fe}}$, respectively, and (ii) assuming that the optical excitation occurs exclusively in Fe following the dominant pump absorption in the Fe layer for Fe-side pumping, see Fig.~\ref{Fig:Abs}. We, furthermore, assume that the electrons in Fe instantaneously reach an electron temperature $T_{e}=500~$K. We obtained this value by fitting a Fermi-Dirac distribution function to the time-dependent photoemission spectra in Fig.~\ref{Fig:DelPESVsEfords} at $\Delta t= 100$~fs. We take material parameters for Au and Fe into account as detailed below. The TTM takes e-ph interaction and diffusive heat transfer into account as described by the coupled differential equations~\cite{bonn_PRB00, Bovensiepen_2007},

\begin{eqnarray}
\label{eq:TTM}
C_{e}(T_{e})\frac{\partial T_{e}}{\partial t}&=& S(\Delta t, z)-g(T_{e}-T_{l}) + \frac{\partial}{\partial z} \left(\kappa
\frac{\partial T_{e}}{\partial z}\right),  \\
C_{l}\frac{\partial T_{l}}{\partial t} &=& g(T_{e}-T_{l}),
\end{eqnarray}
where $C_{e}=\gamma T_{e}$ and $C_{l}$ are the electron and lattice specific heat capacities, respectively, $\kappa$ is the electronic part of the thermal conductivity which is responsible for diffusive electronic transport in the Au layer. The e-ph interaction is modeled by e-ph coupling constants $g$ taken from literature: $g_{\mathrm{Au}} = 2.3\times10^{16}~$Wm$^{-3}$K$^{-1}$~\cite{Brorson1987} and $g_{\mathrm{Fe}} = 9\times10^{17}~$Wm$^{-3}$K$^{-1}$~\cite{kang_2021}. The difference in the e-ph coupling constants is linked to the larger electronic density of states at the Fermi energy in iron compared to gold~\cite{Eiguren_2003}. Diffusive heat transport driven by the gradient $\partial T_{l}/\partial z$ is discarded because it proceeds on time scales $>100$~ps which are not discussed here. The reason is that $T_l$ and in consequence $\partial T_l / \partial z$ are considerably smaller than the respective values for $T_e$. $S(z,t)$ is the source term determined by the energy $C_{e} T_{e}$ deposited in Fe. The electronic thermal conductivity $\kappa$ is taken as temperature dependent~\cite{bonn_PRB00}: $\kappa(T_{e}) = \kappa_{0}T_{e}/T_{l}$. Since the Debye temperature of Au is 165~K, we used the high temperature value of $C_{l}^{\mathrm{Au}}(T \rightarrow\infty)=33$~Jmol$^{-1}$K$^{-1}$ ~\cite{arblaster_2016}. For Fe, whose Debye temperature is 460~K, we take the room temperature value $C_{l}^{\mathrm{Fe}}(T=300~\mathrm{K})=26$~Jmol$^{-1}$K$^{-1}$~\cite{desai_1986}. Using further literature values for Au and Fe $\kappa_{0}^{\mathrm{Au}}=317~$Wm$^{-1}$K$^{-1}$, $\gamma^{\mathrm{Au}}=71$~Jm$^{-3}$K$^{-2}$,  $\kappa_{0}^{\mathrm{Fe}}=80~$Wm$^{-1}$K$^{-1}$, $\gamma^{\mathrm{Fe}}=764$~Jm$^{-3}$K$^{-2}$~\cite{kittel}, the transient electron temperature was calculated. Fig. ~\ref{Fig:TTM}(a-c) depicts the results of this calculation for a gold thickness of $d_{\mathrm{Au}} = 100$~nm which enables us to simultaneously discuss the experimental results of different gold thickness.

With increasing time delay, we observe the cooling of the electron system in Fe driven by e-ph coupling which is very efficient and occurs already around 100~fs, see $z<0$ in Fig.~\ref{Fig:TTM} (a). Simultaneously, electron diffusion is driven by $\partial T_e / \partial z$ across the Fe-Au interface and increases $T_e$ in Au, see $z>0$ in panels (a,b). The cooling of $T_e$ in Au is slower than in Fe because the e-ph coupling in Au is almost 40 times weaker than in Fe. The combination of diffusive electron transport and e-ph coupling determines effective finite time and length scales of diffusive electron transport which increase $T_e$ within 200~fs and 100~nm by approximately 50~K as discussed in the following. Since in Au the electron transport is faster than e-ph coupling, $T_e$ increases on a second, longer time scale for larger distance to the Fe-Au interface and for later delays, see the data for 0.5 and 1.0~ps in panel (b). In agreement with this interplay between e-ph coupling end electron diffusion, panel (c) indicates that the maximum $T_e$ shifts to later $\Delta t$ with increasing $z$. At a distance of 43~nm to the Fe-Au interface the maximum $T_e$ is found at $\Delta t = 190$~fs, while it occurs at 28~nm distance at 110~fs.

A more detailed look suggests to distinguish two regimes of the relaxation dynamics, i.e. the cooling of $T_e$, in Au as a function of $z$. First, for $z \lesssim 20$~nm, i.e. within Au close to the Fe-Au interface, the relaxation in Au is almost as fast as in Fe, which is explained by electron transport from Au back to Fe. This is driven by the faster e-ph coupling in Fe compared to Au which reduces $T_e$ in Fe and results in $\partial T_e / \partial z$ of the opposite sign compared to initial diffusive electron transport into the Au. Within these spatial and temporal regimes diffusive electron transport from Au to Fe occurs and the excess energy in Fe decays too fast for the electrons to couple to phonons in Au and electrons in Au transfer back to Fe and couple to phonons in Fe. This explains why $T_e$ in Au very close to the interface relaxes much faster than in bulk Au~\cite{Hohlfeld2000} and emphasizes that for $z \lesssim 20$~nm the relaxation of $T_e$, and $\epsilon$, is determined by electronic transport effects. Second, for $z > 70$~nm, the electron temperature reaches a maximum in Au at 300~fs and e-ph coupling in Au is a relevant channel for energy dissipation because at these larger $z$ the distance to Fe is too large to compete by diffusive electron transport to Fe with the local e-ph coupling in Au. Third, in between, for 20 nm $< z <$ 70 nm, there is a transition from the low $z$ to the high $z$ regime.

\section{Discussion}
Since the probing occurs at the Au surface for both pump geometries, the experimental results reported above for Au-side and Fe-side pumping allow for a direct comparison of hot electron dynamics that is driven by electrons injected at a defined distance into the Au layer with dynamics that is optically excited at the Au surface. The measured energy distribution of the photo-excited electrons above $E_{\mathrm{F}}$, see Fig.~\ref{Fig:DelPESVsEfords}, exhibits two interesting features for $d_{\mathrm{Au}}$=15, 28~nm. (i) The results for Au-side pumping is characterized by an electron population loss at $E-E_{\mathrm{F}}>0.5$~eV. For Fe-side pumping the population relaxes above 0.5 eV as well, but below 0.3~eV the population increases during the first few 10~fs, in contrast to data for Au-side pumping. (ii) The distribution change with increasing $\Delta t$ indicates a stronger trend towards a thermalized electron distribution for Fe-side than for Au-side pumping. Both observations are in good agreement with the assignment of an excess energy loss of hot electron distributions to transport effects in previous time-resolved photoemission studies on Ru(001)~\cite{liso_APA04}. The hot electrons excited at the Au surface propagate into the depth of the Au film and into Fe which leads to a loss in the time-resolved photoelectron intensity since photoemission is a surface sensitive probe. In case of the Fe-side pumping, the hot electrons scatter with other electrons while they propagate through the whole Au layer and are monitored in the spectrum at a lower electron energy, closer to a thermalized distribution. Nevertheless, the observed electron distributions still feature non-thermal contributions up to $\Delta t = 300$~fs. The case of $d_{\mathrm{Au}}=5$~nm differs characteristically from this behavior because the optical transmission of the pump pulse through the very thin Au layer for Au-side pumping and its absorption in Fe still induces the dominant hot electron fraction, see Sec.~\ref{Sec:Exp}.

This scenario is in good agreement with our analysis of the time-dependent energy density shown in Fig.~\ref{Fig:EdensityVsDeltat}. For Au-side pumping and $d_{\mathrm{Au}}$=15, 28~nm, the excited energy density is completely dissipated within 250~fs while for Fe-side pumping a remnant pedestal at 1~ps is found, that we assign to phonons excited by e-ph coupling. Moreover, in case of Au-side pumping the high-energy fraction of the energy density, i.e. $\epsilon_{0.5}^{1.5}$, accounts for about half of the total $\epsilon_{0.1}^{1.5}$, while  for Fe-side pumping this fraction is with about 20\% much weaker. This behavior confirms the dominant contribution of scattered electrons in case of Fe-side pumping and motivates the comparison of the experimental results for Fe-side pumping with the TTM calculations introduced in Fig.~\ref{Fig:TTM}. In Fig.~\ref{Fig:EdensityVsDeltat} (a-c), we compare the measured energy densities with those determined by the two-temperature model which were calculated by using a thermalized pump-induced change in the population change $\Delta n$ in Eq.~\ref{Eq:EnergyDensity} for the $d_{\mathrm{Au}}$ investigated in the experiment. The temperature of that population change was taken from the two-temperature model calculations and are depicted by the green lines. The relaxation of measured time-dependent energy densities $\epsilon_{0.1}^{1.5}$ follows the modeled transient behavior. We find that the smaller $d_{\mathrm{Au}}$ is, the larger the deviations between experiment and the model become. For  $d_{\mathrm{Au}}=28$~nm, experiment and model match quantitatively. We conclude that for $d_{\mathrm{Au}}=5, 15$~nm the scattering pathway through the Au layer is not sufficient to reach a diffusive limit and we consider that electronic transport processes from Au to Fe, see above, which are faster than the purely diffusive contribution in Eq.~\eqref{eq:TTM}, are responsible for this behavior. In this limit, our observations agree with reports in the literature by Battiato \textit{et al.} that conclude on the importance of super-diffusive transport under similar conditions as discussed here~\cite{Battiato2012} and our own previous work~\cite{beyazit_2020}. For $d_{\mathrm{Au}}=28$~nm, the agreement between experiment and the model suggests that the transport proceeds diffusively and scattering in these thicker films is found to be sufficient to reach this limit.

In panel (d) of Fig.~\ref{Fig:TTM} we replot  the measured $\epsilon_{0.1}^{0.5}$ for all $d_{\mathrm{Au}}$ after multiplication with factors such that the data match at $\Delta t = 0.5$~ps, which works well within the experimental uncertainty. Due to the rather small changes in $T_e$, we discard here the resulting variation in the temperature dependent electronic specific heat and compare the scaled energy density in panel (d) with the calculated electron temperature in panel (c). We find that the experimental and theoretical results agree well qualitatively regarding the coinciding transient behavior for different $d_{\mathrm{Au}}$ at $\Delta t\ge 0.5$~ps. Also, the maxima in $\epsilon_{0.1}^{0.5}(\Delta t, d_{\mathrm{Au}})$ recede to almost half for a change in $d_{\mathrm{Au}}$ from 5 to 28~nm in good agreement with the prediction of the calculation. However, the time delay at which the maxima in $\epsilon_{0.1}^{0.5}(\Delta t, d_{\mathrm{Au}})$ occur for different $d_{\mathrm{Au}}$ show a monotonous increase in $\Delta t$ in case of the calculations. In the experimental results the time delay of the maxima for $\epsilon$ with increasing $d_{\mathrm{Au}}$ appears as non-monotonous. Such deviations from the model behavior could indicate transport effects beyond the diffusive limit for the two thinner layers in agreement with the conclusions above from the energy relaxation. We expect that future experiments will allow more gradual variation in $d_{\mathrm{Au}}$, an improved time resolution to monitor the primary energy injection from Fe to Au, and usage of higher pump fluences to provide improved data quality in order to distinguish diffusive and super-diffusive regimes more rigourously.

\section{Conclusion}
In this work we reported experimental results obtained in linear femtosecond time-resolved photoelectron spectroscopy of epitaxial heterostructures Au/Fe/MgO(001) for different pump excitation geometries. The pump pulse either reaches the Au or the Fe-side of the heterostructure while probing always occurs at the Au surface. We draw conclusions from the measured time-dependent photoelectron emission spectra regarding the transient electron distribution and scattering. In addition, we determined the transient excess energy density from these distributions, which is dissipated to the largest extent within 1~ps for Fe-side pumping and within 300~fs for Au-side pumping. The Fe-side pumping results are compared to a calculation by the TTM which predicts spatio-temporal dynamics under the consideration of e-ph coupling and diffusive electron transport. We conclude that electron transport dynamics proceeds close to a diffusive limit, but at a Au layer thickness of 20 to 30~nm a transition from a super-diffusive to a diffusive regime was identified based on the comparison of the modeled and the measured energy relaxation. Our experimental approach gives clear insight that actual hot electrons propagate through the sample.

\begin{acknowledgments}
Funding by the Deutsche Forschungsgemeinschaft (DFG, German Research Foundation) through Project No. 278162697—SFB 1242 and Project No. BO1823/12 - FOR 5249 (QUAST) is gratefully acknowledged.
\end{acknowledgments}

\providecommand{\noopsort}[1]{}\providecommand{\singleletter}[1]{#1}%

\end{document}